\documentstyle[aps,multicol,psfig,rotating]{revtex}     

\newcommand{\beq}{\begin{equation}}    
\newcommand{\eeq}{\end{equation}}    
\newcommand{\bea}{\begin{eqnarray}}    
\newcommand{\eea}{\end{eqnarray}}    
     
\begin{document}       
\draft
\title{Transition from the Couette-Taylor system to the plane Couette system}  
\author{  
Holger Faisst and Bruno Eckhardt  
}       
\address{Fachbereich Physik, Philipps Universit\"at      
         Marburg, D-35032 Marburg, Germany}      
\maketitle{ }     
    
\begin{abstract}  
We discuss the flow between concentric rotating cylinders  
in the limit of large radii where the system approaches  
plane Couette flow. We discuss how in this limit the   
linear instability that leads to the formation of Taylor  
vortices is lost and how the character of the transition  
approaches that of planar shear flows. In particular,   
a parameter regime is identified where   
fractal distributions of life times and spatiotemporal  
intermittency occur. Experiments in this regime should  
allow to study the characteristics of shear flow  
turbulence in a closed flow geometry.  
\end{abstract}       
    
\begin{multicols}{2}     
The transition to turbulence for a fluid between concentric rotating  
cylinders has attracted much experimental and theoretical  
attention. Ever since Taylor's success \cite{taylor23} in predicting  
and observing the instabilities for the formation of   
vortices the system has become one of the paradigmatic  
examples for the transition to turbulence 
and a large number of bifurcations have been analyzed in considerable detail
\cite{snyder70,marcus84_2,andereck86,nonlin_sc_today}. 
The limiting case of large radii  
and fixed gap width where the effects due to curvature  
become less important and where the system approaches  
plane Couette flow between parallel walls 
has received much less attention.  
In this limit the character of the flow changes:  
plane Couette flow is linearly stable and the  
mechanisms that drive the transition to turbulence  
are still unclear. The question we address here is to what extend the
Couette-Taylor system can be used to
gain insight into the dynamics of plane Couette flow.

This problem is of both experimental and theoretical interest.
As mentioned, the experimental situation for Couette-Taylor flow
is much better, there being numerous facilities and detailed studies
of patterns, boundary effects and critical parameters
\cite{andereck86,nonlin_sc_today,koschmieder93}.
The moving boundaries in plane Couette flow reduce the 
experimental accessibility and 
the possibilities of applying controlled perturbations.
On the theoretical side it is an intriguing question how
the change in stability behaviour from the Couette Taylor
system to the plane Couette system occurs. 
Studies by Nagata \cite{nagata90} show that some  
states from the rotating plane Couette system survive the limiting  
process and appear in finite amplitude saddle node bifurcations 
in the plane Couette system (see also the investigation of this
state by Busse and Clever \cite{busse97}). 
Unless the transition from linear instability 
dominated behaviour in Couette-Taylor flow to 
the shear flow type transition in plane Couette flow 
is singularly connected to the absence of any curvature
it can be expected to happen at a finite radius ratio 
near which interesting dynamical behaviour should occur.

We should mention that there are other useful embeddings of plane
Couette. Busse and Clever \cite{busse97} start from a layer of fluid
heated from below with cross flow and proceed to study the stability
and parameter dependence of the states. And Cherhabili and
Ehrenstein \cite{cherhabili_ehrenstein97}
start from plane Poisseuille flow and find localized
solutions, albeit at Reynolds numbers higher than the ones
studied here.

Our aim here is to follow 
some of the instabilities in the Couette-Taylor system to the limit
of the plane Couette system and to identify the  
parameters where the change in behaviour occurs. In particular, we 
study the transition from laminar Couette flow to Taylor vortices
and the instability of vortices to the formation of wavy vortices.
Note that the asymptotic situation of plane Couette flow can
be characterized by a single parameter, a Reynolds number based
on the velocity difference, whereas Couette Taylor flow has at 
least two parameters, the Reynolds numbers based on the 
velocities of the cylinders. This extra degree of freedom provides
an additional parameter that can be used to modify the flow
without changing the basic features.
 
In cylindrical coordinates $(r,\phi,z)$ the equations of 
motion for the velocity components $(u_r, u_\phi, u_z)$ 
can be written as 
\bea  
\partial_t {u_r} +   
\left( {\bf u} \cdot {\bf \tilde{ \nabla}}  
\right) {u_r}   
-\nu \tilde\Delta {u_r}  +{ \bf \tilde{\nabla} }p   
=\nonumber\\  
\nu\left(\frac{1}{r}\partial_r u_r -\frac{2}{r^{2}}\partial_\phi u_\phi  
 -\frac{1}{r^{2}} u_r \right)+ \frac{1}{r} u_\phi^2  
\label{nst1}  
\\  
\partial_t {u_\phi} +   
\left( {\bf u} \cdot {\bf \tilde{\nabla}}  
\right) {u_\phi}   
-\nu \tilde\Delta {u_\phi}  +{ \bf \tilde{\nabla} }p  
=\nonumber\\  
 \nu\left( \frac{1}{r}\partial_r u_\phi  
+\frac{2}{r^{2}}\partial_\phi u_r -\frac{1}{r^{2}} u_\phi \right)  
- \frac{1}{r}u_ru_\phi   
\\  
\partial_t {u_z} +   
\left( {\bf u} \cdot {\bf \tilde{\nabla}}  
\right) {u_z}   
-\nu \tilde\Delta {u_z}  +{ \bf \tilde{\nabla} }p  
=\nonumber\\     
\nu \frac{1}{r}\partial_r u_z  
\\  
{\bf \tilde{\nabla}}\cdot {\bf u}  
=  
-\frac{1}{r} u_r  
\label{nst4}  
\eea  
where the modified Nabla and Laplace operators are
\bea  
{\bf \tilde{\nabla}}    
&=&   
{\bf e_{r}}    \partial_r +    
{\bf e_{\phi}} \frac{1}{r} \partial_\phi +  
{\bf e_{z}}    \partial_z ,   
\\  
\tilde{\Delta}   
&=&   
\partial_{rr} +   
\frac{1}{r^2}\partial_{\phi\phi} + \partial_{zz}\,, 
\eea  
and where ${\bf e}_i$ are the unit basis vectors \cite{landaulifshitz87}

The terms in eqs. (\ref{nst1})-(\ref{nst4}) are arranged so that all the 
ones on the right hand side vanish when the system approaches 
the plane Couette system, i.e.
in the limit of large radii but finite velocities at the cylinders.   
The remaining ones become the equations of motion for 
plane Couette flow in cartesian coordinates $(x,y,z)$ 
if the identification $x=r$ and 
$y=\phi r$ 
is made. However, there are other ways of taking the limit of a small gap 
that lead to different limiting systems.
For instance, the case of almost corotating cylinders 
with high mean rotation rate gives rise to plane Couette
flow with an additional Coriolis term 
(`rotating plane Couette flow' \cite{nagata90}).
Another limit corresponds to the case of counterrotating 
cylinders with diverging rotation rates \cite{demay92}.
In our numerical work we use the full equations, without any reduction
in terms. This allows us to extend Nagata's work from the 
rotating plane Couette flow to the full Couette-Taylor system. 

The velocities at the 
inner and outer cylinder 
(distinguished by indices $i$ and $o$, respectively) 
are prescribed and define the boundary conditions 
\bea  
u_\phi (r=R_x) &=& \Omega_x R_x\,, \\  
u_r (r=R_x) &=& u_z (r=R_x)=0, \hspace{1cm} x = i,o \,.  
\eea  
For the choice of dimensionless quantities 
we appeal to the plane Couette flow limit. 
There the relevant quantities are the velocity difference   
between the walls, $\Delta U = R_i \Omega_i - R_o \Omega_o$,   
and the gap width $d=R_o-R_i$.   
Without loss of generality we can always assume $\Omega_i \geq 0$.   
The Reynolds number for plane Couette flow   
is based on half the velocity difference and half the   
gap width, 
\beq  
Re = \frac{\Delta U d}{4\nu}\,.  
\eeq  
For the Couette Taylor system there are two Reynolds numbers based
on the gap width and the rotation rates of the inner and outer
cylinders,
\beq
Re_x = R_x\Omega_x d /\nu \,,
\eeq
where the index $x$ can stands for $i$ or $o$, the inner and
outer cylinders. 
The plane Couette flow Reynolds number thus
is $Re = (Re_i-Re_o)/4$. The ratio of these  
Reynolds numbers will be called 
\beq
\tilde \mu=Re_o/Re_i\,,
\eeq
(the tilde is used to distinguish it from 
$\mu=\Omega_o/\Omega_i$, a frequently defined quantity not used here).  
\beq
\eta=R_i/R_o
\eeq
denotes that ratio of radii.
  
Experiments and numerical simulations show that 
plane Couette flow undergoes a subcritical transition to turbulence around  
$Re_{PCF}\approx 320$ \cite{tillmark92,bottin97,se99}. The Couette-Taylor
system shows a first linear instability to the formation
of vortices (Taylor-vortex flow, TVF) at Reynolds numbers that
depend on the rotation rates and the curvature of the cylinders.
In order to see shear flow dominated dynamics the critical
Reynolds number for the linear instability has to be above $Re_{PCF}$.
The formation of TVF occurs at Reynolds numbers 
that can be parametrized in the form  
\beq  
Re = A(\tilde\mu)(1-\eta)^{-1/2} +B(\tilde{\mu})
\label{egform}  
\eeq  
for $\eta \lesssim 1 $ \cite{esser_grossmann96}.  
This number is larger than the   
transitional Reynolds number for plane Couette flow   
if $\eta$ is sufficiently close  
to one. The minimal radius ratio $\eta_{320}$ where the   
linear instability occurs for $Re>320$ strongly depends 
on the ratio of the Reynolds numbers of inner 
and outer cylinder. A few examples for minimal radius  
ratios $\eta_{320}$ are summarized in Table~\ref{tab}. 
  
Very important for the transition to turbulence in linearly
stable systems are nonlinear processes that could 
give rise to some finite amplitude
states, perhaps stationary or periodic, around which the 
turbulent state could form.
One candidate that could serve as a nucleus for 
turbulence in plane Couette flow is the  
stationary state first calculated by Nagata \cite{nagata90}.
He observed that the wavy 
vortices that form in a secondary instability from the  
TVF in the rotating plane Couette system
can be followed to the limit of the  
plane Couette flow where they become part of a saddle 
node bifurcation at finite Reynolds numbers. 
This state was also identified and studied in a different
limiting process by Busse and Clever \cite{busse97}.
They found that the critical axial and azimuthal wavelengths for this state are    
\beq
\lambda_z = \pi$ and $\lambda_{\phi} = 2 \pi\,.
\label{scale}
\eeq   
This is roughly twice the critical wave lengths that
would be expected for Taylor vortices.

We developed a numerical code for the 
solution of the full Navier-Stokes equation using Fourier modes
in axial and azimuthal direction and Legendre collocation in 
the radial direction. The pressure terms were treated by a Lagrange 
method. The period in $z$ and $\phi$ was determined by the 
fundamental wave lengths (\ref{scale}) of
wavy vortex flow.

\begin{figure}    
\begin{center}  
\psfig{file=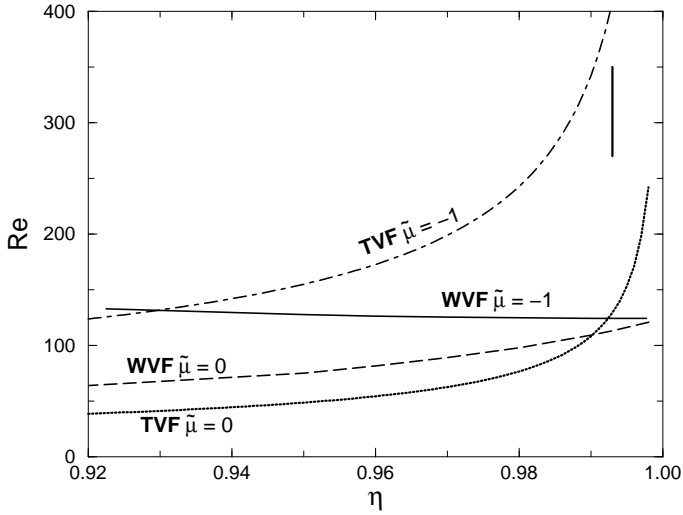,width=90mm}  
\end{center}  
\caption[]{Bifurcations to Taylor vortex flow (TVF) and
wavy vortex flow (WVF) in Couette-Taylor flow   
for the outer cylinder at rest ($\tilde\mu=0$) and   
counter-rotating cylinders ($\tilde\mu=-1$). 
The vertical line indicates the parameter range of the lifetime measurements
of Fig.~\ref{life_times} at $\tilde\mu=-1$.
}\label{instabilities}    
\end{figure}

The continuation of the wavy vortex flow from the Couette-Taylor 
system to the plane Couette system is shown 
in Fig.~\ref{instabilities} for the case of the outer cylinder
at rest ($\tilde\mu=0$) and for counterrotating cylinders with
$\tilde\mu=-1$. For small $\eta$ the wavy vortex develops from
a secondary bifurcation of TVF, but for sufficiently
large $\eta$ the wavy vortex state is created first in a saddle 
node bifurcation. 
The critical Reynolds number for the formation of Taylor vortices
diverges as $\eta$ approaches one, but the one for the formation
of wavy vortices approaches a finite value. Thus the gap
in Reynolds numbers between the two transitions widens and
the region where plane Couette flow like behaviour can be expected
increases with $\eta$ approaching one. The radius ratios $\eta_c$ and 
Reynolds numbers $Re_c$ of
the codimension two point where the instabilities for TVF
and wavy vortex flow cross are listed in 
Table~\ref{tab}.  
The ratio of radii $\eta_c$ where the linear instability of 
Couette flow and of the Taylor vortex flow cross is a non-monotonic
function of the ratio $\tilde \mu$ of rotation speed.
Both the critical Reynolds numbers for the linear
instability of the Couette profile and for the 
formation of wavy vortices increase with decreasing 
$\tilde\mu$, but at different rates and with different
dependencies of $\eta$. As a consequence there seems
to be a local minimum near about $0.93$ for $\tilde\mu$ close to $-1$.

\begin{figure}    
\begin{center}  
\psfig{file=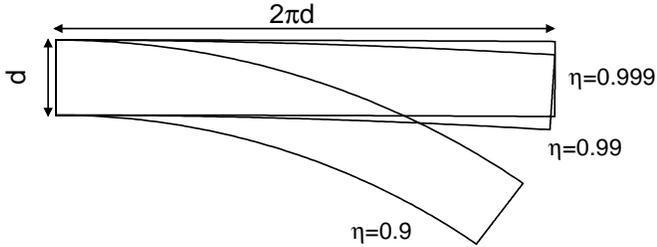,width=90mm}  
\end{center}  
\caption[]{Geometrical curvature of the cylinders in the Couette-Taylor 
flow and  
the plane Couette flow limit. Shown is one fundamental azimuthal wavelength   
for different radius ratios $\eta$ as indicated.   
}\label{geometry}    
\end{figure}

For the parameter value considered here the
curvature of the cylinder walls is geometrically
small (see Fig.~\ref{geometry}). 
On the length of one unit cell in $\phi$-direction the 
relative displacement in radial direction from a planar wall 
is about $\pi(1-\eta)$, i.e. only $3\%$ for $\eta=0.99$.

\begin{figure}    
\begin{center}  
\psfig{file=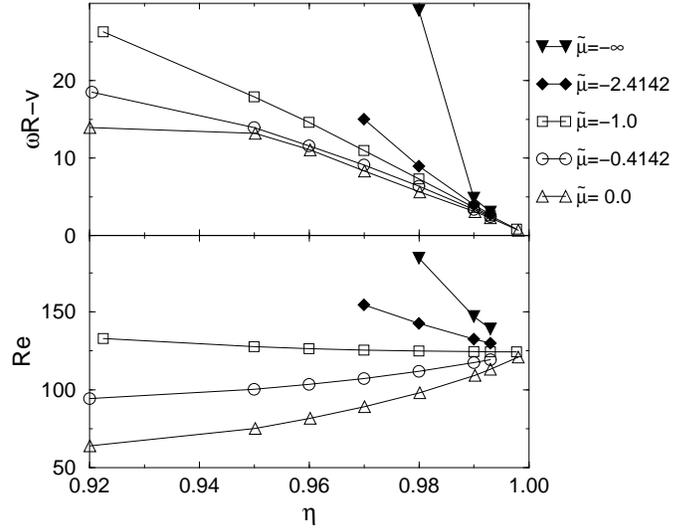,width=90mm}  
\end{center}  
\caption[]{The convergence to the Nagata-Busse-Clever state for
different rotation ratios $\tilde\mu$.  
In the limit of $\eta$ going to one the wavy vortex states  
for all $\tilde\mu$ approach the same flow that moves  
with the mean velocity azimuthally. The top diagram shows  
the rotation speed and the botton one the critical  
Reynolds numbers.  
}\label{NBC1}    
\end{figure}

\begin{figure}    
\begin{center}  
\psfig{file=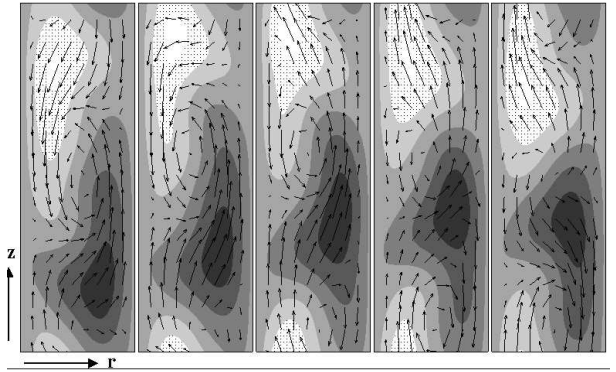,width=80mm }      
\end{center}  
\caption[]{The wavy vortex flow state near the Nagata-Busse-Clever state   
at $\eta=0.993$, $\tilde\mu = -1$ and $Re=124$. Shown is only
the disturbance, without the Couette profile.
The frames from left to right show cuts through the $(r,z)$ plane
at azimuthal wave lengths $\phi=0$, $\pi/4$, $\pi/2$, $3\pi/4$ and $\pi$.   
The vectors indicate the $r$ and $z$ components of the velocity field
and shading the $\phi$-component.  
The inner (outer) cylinder is located at the left (right) 
side of each frame.   
}\label{NBC2}    
\end{figure}

The critical Reynolds number for the formation of 
wavy vortex flow (WVF) seems to converge to the same value for both
ratios $\tilde\mu$ shown in Fig.~\ref{instabilities}. 
The critical Reynolds number
as well as the rotation speed of the wavy vortices for several
different ratios $\tilde \mu$ are collected in
Fig.~\ref{NBC1}. 
The rotation speed is defined as
the angular phase velocity $\omega$ of WVF   
times the mean radius $R=(R_i+R_o)/2$   
minus the mean azimuthal velocity   
$v =(\omega_i R_i+\omega_o R_o)/2$.
For all ratios between the speed of inner and
outer cylinder the critical Reynolds number for the formation of the 
wavy vortex state converges to a value of about 125 and the 
speed of rotation goes to zero. The limiting state 
that is approached
is the stationary Nagata-Busse-Clever state.
The velocity field of a wavy vortex solution at $\eta=0.993$, 
$\tilde\mu=-1$ and $Re=124$ is shown in Fig.~\ref{NBC2};
it differs little from the corresponding
plane Couette state obtained by Busse and Clever \cite{busse97},
both in appearance and in critical Reynolds number.

\begin{figure}  
\begin{center}  
\psfig{file=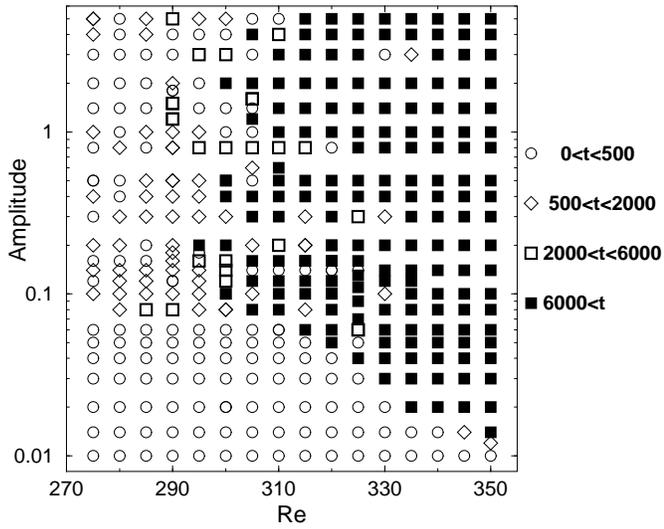,width=90mm}  
\end{center}  
\caption[]{Life time distribution in Couette-Taylor flow  
at $\eta = 0.993, \tilde{\mu} = -1$ and for the indicated range of   
Reynolds numbers.  
}\label{life_times}    
\end{figure}

In the region above the wavy vortex instability but below
the linear instability the dynamics of perturbations shows
the fractal life time pictures familiar from plane
Couette flow \cite{se97}. Fig.~\ref{life_times} shows an example
at a radius ratio of $\eta=0.993$ 
and a Reynolds number ratio of $\tilde{\mu} = -1$. 
The initial state was prepared by
rescaling a WVF field obtained at very low radius ratio and Reynolds number. 
It is interesting to
note that even with this initial condition, which is at least
topologically close to the Nagata-Busse-Clever state,
it is not possible to realize a turbulent signal in its 
neighborhood: the state quickly leaves this region in phase space.
One might have hoped that in spite of the linear instability
of the Nagata-Busse-Clever state other states created
out of secondary bifurcations could have supported some
turbulent dynamics in its neighborhood, but the numerical
experiments do not support this.
The gap between the Reynolds number where the WVF state is
formed and the one where typical initial conditions 
become turbulent is about the same as in plane Couette flow:
the WVF states forms around $Re=125$ and the transition to
turbulence, based on the requirement that half of all 
perturbations induce a long living turbulent state, occurs
near a value of $Re_{trans} = 310$, very much as in 
plane Couette flow \cite{se99}.
  
In summary, we have identified parameter ranges in the    
Couette-Taylor system
where some of the characteristics of the plane   
Couette system can be found. 
These parameter ranges include radius ratios that can be realized 
experimentally. 
Investigations in this 
regime should be rewarding as they open up the  
possibility to study the properties of the transition 
in a closed geometry and to switch continuously between supercritical 
and subcritical transition to turbulence. 
The observation of  
a codimension two point where the linear instability to 
TVF and the secondary instability to  
wavy vortex flow cross should provide a starting point for  
further modelling of the transition in terms of 
amplitude equations. 
\subsection*{Acknowledgments}
This work was financially supported by the Deutsche Forschungsgemeinschaft.

  
\narrowtext    

\begin{table}  
\begin{tabular}{l| c | c | c | c | c }
$\tilde{\mu}$&  $A$ & $B$  & $\eta_{320}$ &$\eta_c$ & $Re_c$   \\\hline
\ 0.0        & 10.8 & 0.5 & 0.999 & 0.990    &   109           \\
-0.4142      & 16.8 & 0.8 & 0.997 & 0.977    &   110          \\
-1.0         & 33.9 & 3.2 & 0.989 & 0.929    &   131          \\
-2.4142      & 53.6 & 4.6 & 0.971 & $\approx$ 0.94 & $\approx$ 220 
\end{tabular}
\caption[]{Parameters connected with the Couette-Taylor system
in the limit of large radii. $A$ and $B$ are
the coefficients in the parametrization (\ref{egform}) of the primary instability.
$\eta_{320}$ is the radius ratio where the primary instability lies
above $Re = 320$; finally,   
            $\eta_c$ and $Re_c$ are the parameter values for the 
	    crossing of the stability curves for Taylor vortex flow
	    and wavy vortex flow.
}  
\label{tab}  
\end{table}  

\end{multicols}  

\begin{references}    
 
\bibitem{taylor23}
G.I. Taylor.
\newblock {\em Phil. Trans. Roy. Soc. London}, {\bf 223}(A), 289, (1923).

\bibitem{snyder70}
H.A. Snyder.
\newblock {\em Int. J. Non-Linear Mechanics}, {\bf 5}, 659, (1970).

\bibitem{marcus84_2}
P.S. Marcus.
\newblock {\em J. Fluid Mech.}, {\bf 146}, 65, (1984).

\bibitem{andereck86}
C.D. Andereck{,} S.S. Liu{,}~H.L. Swinney.
\newblock {\em J. Fluid Mech.}, {\bf 164}, 155, (1986).

\bibitem{nonlin_sc_today}
R.~Tagg.
\newblock {\em Nonlinear Science Today}, {\bf 4}(3), (1994).

\bibitem{koschmieder93}
L.~Koschmieder.
\newblock Cambridge University Press, (1993).
  
\bibitem{nagata90}  
M.~Nagata.  
\newblock {\em J. Fluid Mech.}, {\bf 217}, 519, (1990).    
  
\bibitem{busse97}  
R.M. Clever{,}~F.H. Busse.  
\newblock {\em J. Fluid Mech.}, {\bf 344}, 137, (1997).  


\bibitem{cherhabili_ehrenstein97}
A.~Cherhabili{,}~U. Ehrenstein.
\newblock {\em J. Fluid Mech.}, {\bf 342}, 159, (1997).


            
\bibitem{landaulifshitz87}
L.D. Landau{,}~E.M. Lifshitz.
\newblock Pergamon Press, (1987).    

\bibitem{demay92}  
Y.~Demay{,} G.~Iooss{,} P.~Laure.  
\newblock {\em Eur. J. Mech., B/Fluids}, {\bf 11}(5), 621, (1992).  
   
\bibitem{tillmark92}  
N.~Tillmark{,}~P.H. Alfredsson.  
\newblock {\em J. Fluid Mech.}, {\bf 235}, 89, (1992).      
 
\bibitem{bottin97}   
S.~Bottin{,} O. Dauchot{,}~F. Daviaud.  
\newblock {\em Phys. Rev. Lett.}, {\bf 79}(22), 4377, (1997).         

\bibitem{se99}  
A.~Schmiegel{,} B.~Eckhardt.  
\newblock Dynamics of perturbations in plane {C}ouette flow.  
\newblock {\em submitted}.     
 
\bibitem{esser_grossmann96}  
A.~Esser{,} S.~Grossmann.  
\newblock {\em Phys. Fluids}, {\bf 8}(7), 1814, (1996).   
  
\bibitem{se97}  
A.~Schmiegel{,} B.~Eckhardt.  
\newblock {\em Phys. Rev. Lett.}, {\bf 79}(26), 5250, (1997).     
  
\end{references}
\end{document}